\documentclass[sigconf]{acmart}

\usepackage{xspace}
\usepackage[capitalize,noabbrev,nameinlink]{cleveref}
\usepackage{hyperref}  
\usepackage{diagbox}
\usepackage{multirow}
\usepackage{enumitem}
\usepackage{balance}
\DeclareMathOperator*{\argmax}{arg\,max}

\begin{document}
\copyrightyear{2025}
\acmYear{2025}
\setcopyright{acmlicensed}\acmConference[SIGIR '25]{Proceedings of the 48th International ACM SIGIR Conference on Research and Development in Information Retrieval}{July 13--18, 2025}{Padua, Italy}
\acmBooktitle{Proceedings of the 48th International ACM SIGIR Conference on Research and Development in Information Retrieval (SIGIR '25), July 13--18, 2025, Padua, Italy}
\acmDOI{10.1145/3726302.3731961}
\acmISBN{979-8-4007-1592-1/2025/07}
%%
%% The "title" command has an optional parameter,
%% allowing the author to define a "short title" to be used in page headers.
\title{Learning Universal User Representations Leveraging Cross-domain User Intent at Snapchat}

%%
%% The "author" command and its associated commands are used to define
%% the authors and their affiliations.
%% Of note is the shared affiliation of the first two authors, and the
%% "authornote" and "authornotemark" commands
%% used to denote shared contribution to the research.
% \author{Clark Mingxuan Ju, Leonardo Neves, Bhuvesh Kumar, Liam Collins, Tong Zhao, Yuwei Qiu, Ching Dou, Sohail Nizam, Sen Yang, Neil Shah}
% \email{mju,lneves,bkumar4,lcollins2,tong,yqiu,snizam,qdou,syang3,nshah@snap.com}
% \affiliation{%
%   \institution{Snap Inc.}
%   \city{Santa Monica}
%   \state{California}
%   \country{USA}
% }

% Combined author block
% \author{%
%   Clark Mingxuan Ju, Leonardo Neves, Bhuvesh Kumar, Liam Collins, Tong Zhao, Yuwei Qiu, Ching Dou, Yang Zhou, Sohail Nizam, Rengim Ozturk, Yvette Liu, Sen Yang, Manish Malik, Neil Shah%
% }
% \affiliation{%
%   \institution{Snap Inc.} %
%   % \city{Bellevue, Santa Monica, Palo Alto, San Francisco} %
%   \country{Santa Monica, CA, USA} %
% }
% \email{{mju, lneves, bkumar4, lcollins2, tzhao, yqiu, qdou, yzhou5, snizam, rozturk, yliu7, syang3, mmalik, nshah}@snap.com}

\author{Clark Mingxuan Ju}
\affiliation{
  \institution{Snap Inc.}
  \city{Bellevue}
  \country{USA}
}
\email{mju@snap.com}

\author{Leonardo Neves}
\affiliation{
  \institution{Snap Inc.}
  \city{Santa Monica}
  \country{USA}
}
\email{lneves@snap.com}

\author{Bhuvesh Kumar}
\affiliation{
  \institution{Snap Inc.}
  \city{Bellevue}
  \country{USA}
}
\email{bkumar4@snap.com}

\author{Liam Collins}
\affiliation{
  \institution{Snap Inc.}
  \city{Bellevue}
  \country{USA}
}
\email{lcollins2@snap.com}

\author{Tong Zhao}
\affiliation{
  \institution{Snap Inc.}
  \city{Bellevue}
  \country{USA}
}
\email{tzhao@snap.com}

\author{Yuwei Qiu}
\affiliation{
  \institution{Snap Inc.}
  \city{Palo Alto}
  \country{USA}
}
\email{yqiu@snap.com}

\author{Qing Dou}
\affiliation{
  \institution{Snap Inc.}
  \city{Palo Alto}
  \country{USA}
}
\email{qdou@snap.com}

\author{Yang Zhou}
\affiliation{
  \institution{Snap Inc.}
  \city{San Francisco}
  \country{USA}
}
\email{yzhou5@snap.com}

\author{Sohail Nizam}
\affiliation{
  \institution{Snap Inc.}
  \city{Santa Monica}
  \country{USA}
}
\email{snizam@snap.com}

\author{Rengim Aykan Ozturk}
\affiliation{
  \institution{Snap Inc.}
  \city{Santa Monica}
  \country{USA}
}
\email{rozturk@snap.com}

\author{Yvette Liu}
\affiliation{
  \institution{Snap Inc.}
  \city{Bellevue}
  \country{USA}
}
\email{yliu7@snap.com}

\author{Sen Yang}
\affiliation{
  \institution{Snap Inc.}
  \city{Santa Monica}
  \country{USA}
}
\email{syang3@snap.com}

\author{Manish Malik}
\affiliation{
  \institution{Snap Inc.}
  \city{Palo Alto}
  \country{USA}
}
\email{mmalik@snap.com}

\author{Neil Shah}
\affiliation{
  \institution{Snap Inc.}
  \city{Bellevue}
  \country{USA}
}
\email{nshah@snap.com}

\renewcommand{\shortauthors}{Clark Mingxuan Ju, et al.}
\begin{CCSXML}
<ccs2012>
   <concept>
       <concept_id>10002951.10003317.10003331.10003271</concept_id>
       <concept_desc>Information systems~Personalization</concept_desc>
       <concept_significance>500</concept_significance>
       </concept>
 </ccs2012>
\end{CCSXML}

\ccsdesc[500]{Information systems~Personalization}
%%
%% The abstract is a short summary of the work to be presented in the
%% article.
\begin{abstract} 

The development of powerful user representations is a key factor in the success of recommender systems (RecSys).
Online platforms employ a range of RecSys techniques to personalize user experience across diverse in-app surfaces.
User representations are often learned individually through user's historical interactions within each surface and user representations across different surfaces can be shared post-hoc as auxiliary features or additional retrieval sources. 
While effective, such schemes cannot directly encode collaborative filtering signals \emph{across} different surfaces, hindering its capacity to discover complex relationships between user behaviors and preferences across the whole platform.
To bridge this gap at Snapchat, we seek to conduct universal user modeling (UUM) across different in-app surfaces, learning general-purpose user representations which encode behaviors across surfaces.
Instead of replacing domain-specific representations, UUM representations capture \emph{cross-domain} trends, enriching existing representations with complementary information.
This work discusses our efforts in developing initial UUM versions, practical challenges, technical choices and modeling and research directions with promising offline performance. 
Following successful A/B testing, UUM representations have been launched in production, powering multiple use cases and demonstrating their value. 
UUM embedding has been incorporated into \textit{(i) Long-form Video embedding-based retrieval}, leading to \textbf{2.78\%} increase in Long-form Video Open Rate, \textit{(ii) Long-form Video L2 ranking}, with \textbf{19.2\%} increase in Long-form Video View Time sum, \textit{(iii) Lens L2 ranking}, leading to \textbf{1.76\%} increase in Lens play time, and \textit{(iv) Notification L2 ranking}, with \textbf{0.87\%} increase in Notification Open Rate.

\end{abstract}

%%
%% The code below is generated by the tool at http://dl.acm.org/ccs.cfm.
%% Please copy and paste the code instead of the example below.
%%

%%
%% Keywords. The author(s) should pick words that accurately describe
%% the work being presented. Separate the keywords with commas.
\keywords{Recommender Systems, Universal User Modeling, Sequential Recommendation, Cross-domain Recommendation}
%% A "teaser" image appears between the author and affiliation
%% information and the body of the document, and typically spans the
%% page.

%%
%% This command processes the author and affiliation and title
%% information and builds the first part of the formatted document.
\maketitle
\section{Introduction}
Recommendation systems (RecSys) are fundamental to the success of modern social media applications~\citep{pal2020pinnersage,gomez2015netflix,van2013deep,schafer1999recommender}, driving user engagement across diverse in-app surfaces like Snapchat's various features~\citep{shi2023embedding,tang2022friend,sankar2021graph,kolodner2024robust,kung2024improving}.
A crucial component of effective RecSys is the learning of user representations, which encode users' interaction histories into low-dimensional vectors~\citep{rendle2009bpr,koren2009matrix}.
User representations are usually trained within each individual domain (surface), with cross-domain information sharing limited to post-hoc strategies like employing representations from one domain as auxiliary features for another. 
Leveraging shared user behaviors across domains can improve representation learning, especially for users with sparse interactions in individual domains~\cite{zhu2021cross,zhu2024modeling,park2024pacer}.

While effective, post-hoc approaches overlook the potential for directly leveraging collaborative filtering signals \emph{across} different domains, thereby limiting the ability to jointly capture platform-wide relationships between user behavior and preferences~\citep{ma2019pi,lin2024mixed,park2023cracking}.
Furthermore, this approach does not scale well when multiple domains co-exist -- in order to capture cross-domain signals across multiple surfaces, one needs to aggregate user embeddings from each individual domain.
This fragmented approach creates complex dependencies between domains, not only necessitating costly operational overhead and  cross-team coordination efforts, but also placing a significant burden on infrastructure and resources.

In light of this, we propose to conduct universal user modeling (UUM).
UUM is a general learning paradigm that leverages user behaviors from multiple in-app surfaces and produces user-level embeddings that capture collaborative signals across different domains. 
As a joint work between researchers and engineers at Snapchat, inspired by our previous work~\citep{fang2024general}, this manuscript explores modeling choices in developing initial UUM versions that already have produced strong post-launch gains and proposes future directions with promising offline performance. 

\section{Preliminary}

% \noindent This work considers CDSR where user sequences can contain an arbitrary number of domains.
We formulate UUM as a cross-domain sequential recommendation problem, where user representations should be predicative of future user behaviors across multiple domains. 
Formally, we consider a set of domains $\mathcal{D} = \{d_1, d_2, ...\}$ where $|\mathcal{D}| \geq 2$.
For each domain $d \in \mathcal{D}$, we represent a user's interaction history within that domain as a sequence $X^d = [x^d_1, x^d_2, x^d_3,... ]$. Here $x^d_m \in \mathcal{V}^d$ is the $m$-th item the user has interacted with in domain $d$ and $\mathcal{V}^d$ refers to all items in domain $d$.
The user's overall interaction history across all domains is represented by the sequence $X = [x_1, x_2, x_3,..., x_M]$, where $x_m \in \mathcal{V}$ can come from any domain $d \in \mathcal{D}$, $\mathcal{V}$ denotes the set of all items in all domains, and $M$ refers to the sequence length.
% \footnote{For the ease of notation and reading, here we assume all user sequences have the same length, which is data-dependent and not always true in implementation.}.
$X$ can be generated by stitching $\{X^d\}_{d \in \mathcal{D}}$ together in a way s.t. the user interacted with $x_{m-1}$ earlier than $x_m$ and $|X|=M=\sum_{d \in \mathcal{D}}|X^d|$.
As an initial step, we formulate UUM training as a retrieval task, where the model maximizes the probability of retrieving the next item that a user will interact with, formulated as:
$
    \argmax_{x^* \in \mathcal{V}} P(x^* | X, \{X^d\}_{d \in \mathcal{D}}),
    \label{prelim:objective}
$
where $x^*$ refers to the next item the user will interact with.
\begin{figure}
    \centering
    \includegraphics[width=0.48\textwidth]{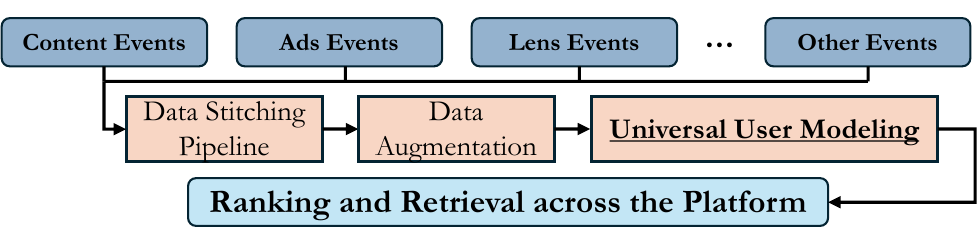}
    \vspace{-0.3in}
    \caption{Overview of universal user modeling (UUM).
    % Data stitching pipeline merges data from multiple in-app surfaces. UUM model is trained on merged sequential data and generates general-purpose user representations, which can be used for either retrieval or ranking in various applications. 
    }
    \label{fig:system}
    \vspace{-0.1in}
\end{figure}
\section{Method Overview}
\begin{figure}
    \centering
    \includegraphics[width=0.48\textwidth]{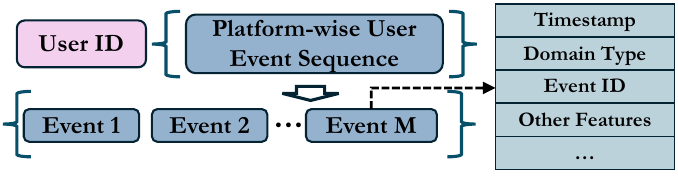}
    \vspace{-0.3in}
    \caption{Each row of UUM data given an user. 
    }
    \label{fig:raw_data}
    \vspace{-0.1in}
\end{figure}

% \subsection{Data Preparation}
\textbf{Raw Data Generation.}
The raw data for UUM includes user events coming from multiple domains including but not limited to Content (Long-form Videos), Ads, and Lens engagement in Snapchat.
We tabularize the data s.t. each row is keyed by user ID, as shown in \cref{fig:raw_data}.
Each column (except the user ID) represents a feature, with values sorted according to a specific ordering (e.g., event IDs by timestamp). 
For features with sparsity across the user base, the corresponding columns are padded with null values at indices where feature values are undefined (e.g., vendor ID for Ads is undefined for Content).
We restrict the maximum sequence length to 5,000 to prevent a small number of users with long sequences.
In the user sequence, we keep high-intent events as many as possible and trim low-intent events based on the timestamp. 

\noindent \textbf{Data Augmentation.}
The number of user interactions varies greatly due to the power-law distribution common in social data~\citep{muchnik2013origins}. 
To enable efficient batched training, shorter sequences must be padded to match the longest sequence in a batch. 
This padding, however, significantly increases computational overhead.  
Moreover, using only a single training example per user may be insufficient for training complex sequential models, which typically require large amounts of data~\citep{zhai2024actions}.
Therefore, to increase the number of training examples and reduce the impact of padding, we use a sliding window approach to divide each user's full user sequence into multiple shorter subsequences, each of which serves as an independent training sample.
We divide the full sequence into multiple subsequences, each of which has length 800. 
The decision to use 800-length subsequences reflects a balance between capturing long-range dependencies and the practical limitations of computational resources, a compromise supported by empirical results.

\subsection{Universal User Modeling}
\begin{figure}
    \centering
    \includegraphics[width=0.48\textwidth]{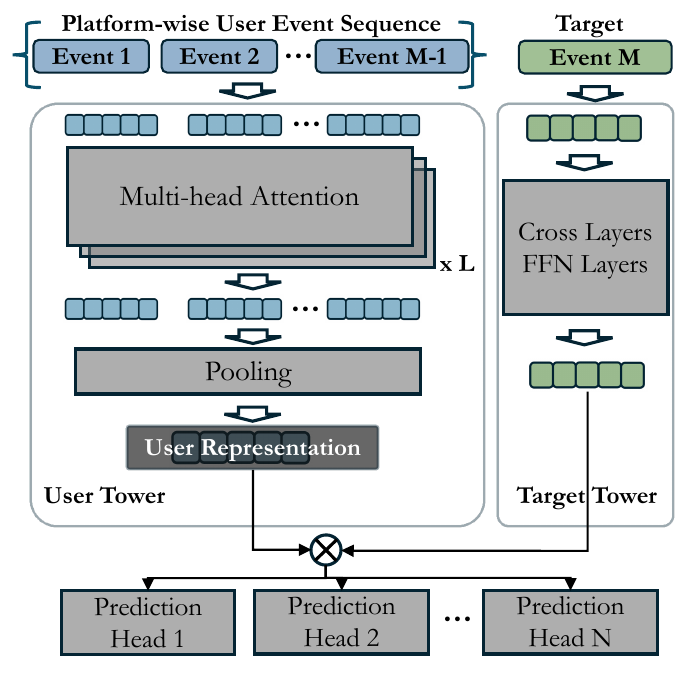}
    \vspace{-0.3in}
    \caption{
    Model architecture of UUM. 
    % The current UUM model follows a two-tower architecture. The user tower consumes user behavioral sequences from multiple domains using multi-head attention and the final user representation is generated by pooling from latent representations of all behaviors.
    % The target tower takes the last N events in the sequence and converts it into a target representation to fulfill the multi-task training objective.
    }
    \label{fig:model}
    \vspace{-0.1in}
\end{figure}
\subsubsection{Model Architecture}
The overall model architecture is shown in \cref{fig:model}. We explore a model architecture based on self-attention in transformer~\citep{kang2018self}, owing to its strong capabilities of modeling sequences that have been well-demonstrated in other fields~\citep{vaswani2017attention}.
Formally, given a sequence $X$, we first convert each event $x_i \in X$ into a latent representation, formulated as:
\begin{equation}
    \mathbf{h}_i = \text{FFN}(\text{Concat}(\text{Feature}(x_i), d(x_i))),
    \label{eq:feature_prep}
\end{equation} 
where $\mathbf{h}_i \in \mathbb{R}^f$ refers to the $f$-dimensional latent representation for $x_i$.
Here, $\text{Feature}(\cdot)$ maps the raw features of $x_i$ into an embedding.
This may involve operations like looking up ID embeddings for categorical features and concatenating the results. 
$d(x_i)$ denotes the domain for $x_i$, $\text{Concat}(\cdot)$ refers to the concatenation operation, and $\text{FFN}(\cdot, \cdot)$ is a function that projects feature vectors from different domains into a shared latent space.

With the latent representation $\mathbf{H} \in \mathbb{R}^{M \times f}$ for $X$, we feed $\mathbf{H}$ through stacked $L$ multi-head attention layers\footnote{We omitted formulation for multi-head attention due to space limitation.} and pool embeddings of all tokens to constitute the final UUM embedding as:
\begin{equation}
    \mathbf{h}^* = \text{Pooling}(\text{MHA}(\mathbf{H})),
\end{equation}
where $\mathbf{h}^* \in \mathbb{R}^f$ denotes the UUM embedding for user with sequence $X. $In our case, we explore a weighted summation as our pooling. 
Other selections of pooling functions will be iterated in the future. 

\subsubsection{Training Objectives}
To train the model, we regard the last event in the sequence as the label event and generate the user representation using all remaining prior events. 
We process the last event the same way as described in \cref{eq:feature_prep} and further pass the latent representation through a series of cross layers and feed-forward neural networks to generate the target representation. 
We combine the target representation and user representation using an element-wise summation operation and feed the merged pair-wise representation to different prediction heads for optimization. 
The model is trained on a multi-task objective consisting of a next event prediction task formulated as sample softmax with in-batch negatives~\citep{wu2024effectiveness} and next event property prediction tasks formulated as either cross-entropy or mean squared error.

% \subsection{Offline Evaluation}
% To iterate over modeling choices, we conduct offline evaluation on our Content L1 ranking model.
% The base model we compare different variants against is the ranking performance without the incorporation of UUM representation, denoted as no-uum.
% In this setup, 

\subsection{A/B Testing and Launch Impact}
Since its initial launch at Snapchat, UUM has demonstrated promising improvements in retrieval and ranking performance across multiple applications, resulting in significantly increased user engagement. 
Use cases include but are not limited to the following:

\vspace{0.05in}
\noindent \textbf{Long-form Video Embedding-based Retrieval (EBR)} \\
\noindent We leveraged UUM embeddings to augment our embedding-based retrieval system for long-form videos, incorporating them as auxiliary features. The results, presented in \cref{tab:ab_ebr}, demonstrate the positive impact of this approach, with a 2.78\% improvement in long-form video open rates and a 0.04\% increase in daily active users, affirming the value of UUM embeddings for enhancing user engagement throughout the recommendation process.

% Snapchat's Long-form video EBR system originally used real-time user historical features for personalized ranking.  However, the cost of fetching and maintaining these features introduced significant latency. To address this, we replaced the real-time features with UUM embeddings.  
% This optimization resulted in a 45\% reduction in EBR online inference latency.  Importantly, this change did not negatively impact ranking performance.  
% In fact, the UUM-based EBR performed as well as, or better than, the original system, as evidenced by the results presented in \cref{tab:ab_ebr}.

\vspace{0.05in}
\noindent \textbf{Long-form Video L2 Ranking} \\
\noindent Besides retrieval, we also applied UUM embeddings to L2 rankers for Long-form videos.
As shown in \cref{tab:ab_l2}, the incorporation of UUM embedding resulted in improvements across several key metrics (e.g., 19.2\% improvement to Long-form Video view time), demonstrating the effectiveness of UUM embeddings for multiple stages within the recommendation pipeline.

\vspace{0.05in}
\noindent \textbf{Lens L2 Ranking} \\
\noindent Furthermore, we integrated UUM embeddings into the L2 ranking model for Snapchat Lenses.  This application of UUM embeddings also yielded positive results, enhancing the relevance and engagement of recommended Lenses, as detailed in \cref{tab:ab_lens}. This confirms the versatility of UUM embeddings across different content types and recommendation tasks.
\begin{table}
    \centering
    \caption{Improvements (\%) to A/B metrics in Long-form Video embedding-based retrieval brought by UUM representations.}
    \vspace{-0.1in}
    \begin{tabular}{l|c}
    \toprule 
    Metrics &  Improvements (\%)\\ 
    \midrule
    %Online Inference Latency &  -45\%  \\
    Long-form Video Open Rate &  +2.78\%  \\
    Long-form Video Direct Open Sum & +2.86\% \\ %$\pm$ 0.31\% \\
    \midrule
    \multicolumn{2}{c}{User Engagement} \\ 
    \midrule
    App Open Daily Active Users & +0.04\% \\ %$\pm$ 0.02\% \\
    Content View Daily Active Users & +0.08\% \\% $\pm$ 0.03\% \\
    \bottomrule
    \end{tabular}
    \label{tab:ab_ebr}
\end{table}

\begin{table}
    \centering
    \caption{Improvements (\%) to A/B metrics in Long-form Video L2 Ranking brought by UUM representations.}
    \vspace{-0.1in}
    \begin{tabular}{l|c}
    \toprule 
    Metrics &  Improvements (\%)\\ 
    \midrule
    Long-form Video View Time Sum & +19.20\% \\ %$\pm$ 0.59\% \\
    Long-form Video Direct Open Sum & +1.01\% \\ % $\pm$ 0.27\% \\ 
    \midrule
    \multicolumn{2}{c}{User Engagement} \\ 
    \midrule
    Long-form Video View Time & +0.28\%\\ %$\pm$ 0.11\% \\
    \bottomrule
    \end{tabular}
    \label{tab:ab_l2}
\end{table}

\begin{table}
    \centering
    \caption{Improvements (\%) to A/B metrics in Lens L2 Ranking brought by UUM representations.}
    \vspace{-0.1in}
    \begin{tabular}{l|c}
    \toprule 
    Metrics &  Improvements (\%)\\ 
    \midrule
    Lens Long Play & +1.76\% \\%$\pm$ 0.23\% \\
    Lens Swipe & +0.72\% \\ % ±0.20\% \\
    % \midrule
    % \multicolumn{2}{c}{User Engagement} \\ 
    % \midrule
    % Content View Time & 0.28\% $\pm$ 0.11\% \\
    \bottomrule
    \end{tabular}
    \label{tab:ab_lens}
    \vspace{-0.1in}
\end{table}

\begin{table}
    \centering
    \vspace{-0.05in}
    \caption{Improvements (\%) to A/B metrics in Notification L2 Ranking brought by UUM representations.}
    \vspace{-0.1in}
    \begin{tabular}{l|c}
    \toprule 
    Metrics &  Improvements (\%)\\ 
    \midrule
    Notification Open Rate & +0.87\% \\ % ±0.09\% \\
    Notification Sent Volume & +0.45\% \\ % ±0.04\% \\
    % \midrule
    % \multicolumn{2}{c}{User Engagement} \\ 
    % \midrule
    % Content View Time & 0.28\% $\pm$ 0.11\% \\
    \bottomrule
    \end{tabular}
    \label{tab:ab_notif}
    \vspace{-0.1in}
\end{table}

\vspace{0.05in}
\noindent \textbf{Notification L2 Ranking} \\
\noindent The benefits of UUM extended to our notification system as well.  By incorporating UUM embeddings into the L2 ranking model for notifications, we improved the relevance and timeliness of the notifications users received. This led to increased notification engagement, as shown in \cref{tab:ab_notif}. This demonstrates UUM's effectiveness in optimizing user experience beyond content feeds.

\begin{figure}[t]
    \centering
    % \vspace{-0.1in}
    \includegraphics[width=0.5\textwidth]{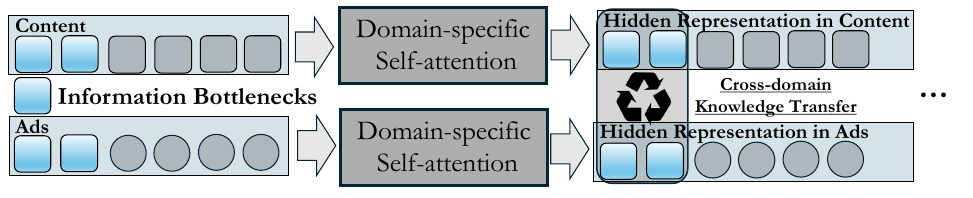}
    \vspace{-0.3in}
    \caption{
    Domain-specific attention with IB. 
    % Cross-domain communication only happens through IB tokens, which enforces structured information transfer by designating a small set of dedicated tokens as attention bottlenecks.
    % Only two domains are drawn for demonstration purposes and this scheme is applicable to multiple domains.
    }
    \label{fig:ib}
    \vspace{-0.1in}
\end{figure}
\subsection{Modeling Choices under Iteration}
Leveraging the strong foundation of UUM's success, we are concurrently investigating advanced modeling techniques to drive continued performance improvements. 
This section details promising avenues we have pursued, which have resulted in positive offline metric results and are awaiting A/B testing validation.

\begin{figure}[t]
    \centering
    \includegraphics[width=0.48\textwidth]{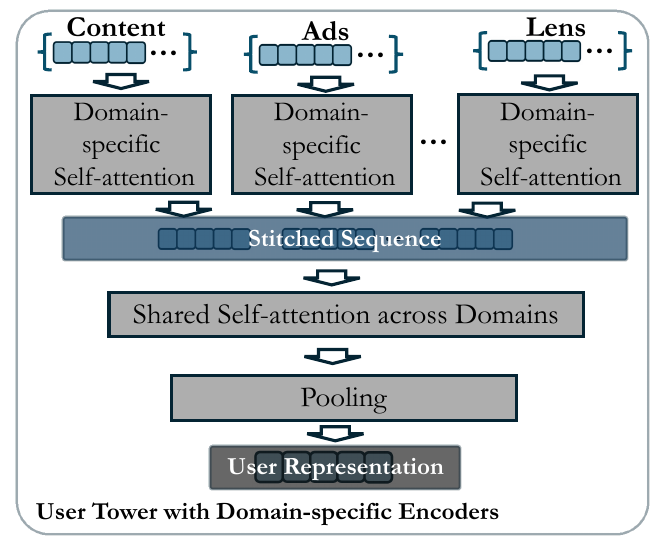}
    \vspace{-0.2in}
    \caption{
    User tower in UUM with domain-specific encoders.
    % Individual subsequences in each domain will be first consumed by domain-specific self-attention encoders first to capture domain-specific sequentiality. 
    % Then we stitch latent representations in all domains chronologically and use a shared encoder to generate the final UUM representation. 
    }
    \label{fig:domain_encoder}
\end{figure}

\subsubsection{Domain-specific Encoders}
While a shared encoder effectively integrates cross-domain event sequences and captures collaborative information, it rests on a critical assumption: that user behaviors across domains are drawn from a uniform distribution and can be accurately represented in a shared feature space. 
However, real-world cross-domain data often exhibits imbalances, leading to asymmetric impacts across domains. 
This can introduce noise, potentially misleading the learning model, rather than simply augmenting knowledge~\citep{park2024pacer,zhu2024modeling,park2023cracking}.
To address the potential for noise and asymmetric impact caused by imbalanced cross-domain data, we introduce a refined UUM framework. 
This framework aims to reconcile the benefits of shared representation with the preservation of domain-specific information. 
Specifically, as depicted in \cref{fig:domain_encoder}, we utilize independent domain encoders to capture intra-domain sequentiality and collaborative information. 
These learned representations are then unified through a shared self-attention mechanism, enabling the model to discern and leverage genuine cross-domain collaborative patterns while mitigating the effects of imbalance.
Thus, we achieve an equilibrium between shared learning and domain-specific representation, addressing the inherent challenges of imbalanced cross-domain data.

\subsubsection{Self-attention with Information Bottlenecks}

Inspired by Information Bottleneck (IB) token research for knowledge transfer~\citep{nagrani2021attention,tsai2019multimodal}, we introduce IB tokens into multi-domain sequences, as shown in \cref{fig:ib}.
These tokens act as attention bottlenecks, restricting cross-domain interaction. Each domain processes its sequence and its IB token separately. 
IB tokens attend to their respective domain event, capturing domain-specific knowledge. 
Cross-domain exchange occurs solely via layer-normalized element-wise summation of IB tokens. Subsequently, domain elements re-attend to the updated IB tokens to incorporate transferred knowledge \cite{ju2025revisiting}.

\begin{table}
    \centering
    \vspace{-0.05in}
    \caption{Offline evaluation (Recall@20 and NDCG@20) on modeling choices we are iterating over. }
    \vspace{-0.15in}
    \begin{tabular}{l|cc}
    \toprule 
    Model Variant &  Recall@20 & NDCG@20\\ 
    \midrule
    Base &  0.483 & 0.254  \\
    \;\; + Domain-specific Encoder & 0.519 & 0.270 \\
    \;\; + Information Bottleneck     & 0.522 & 0.281 \\
    \bottomrule
    \end{tabular}
    \label{tab:offline}
    \vspace{-0.2in}
\end{table}

\subsubsection{Offline Evaluation}

To approximate the effectiveness of these two modeling choices, we train the UUM model with these choices the same way as demonstrated in \cref{fig:model} and evaluated these two modeling choices on the task of next-event retrieval. 
For each testing example, we randomly select 50,000 random negatives from all events to calculate retrieval metrics (i.e., Recall@20 and NDCG@20). Offline performance is shown in \cref{tab:offline}.
We observe that both domain-specific encoders and information bottleneck tokens lead to improvements in Recall@20 and NDCG@20 compared to the base UUM model, demonstrating the efficacy of these techniques in addressing the challenges of imbalanced cross-domain data.

\section{Conclusion}
This paper introduced Universal User Modeling (UUM), a paradigm for learning general-purpose user representations across Snapchat's diverse in-app surfaces. By aggregating user behaviors from multiple surfaces and modeling this as a cross-domain sequential recommendation problem, UUM effectively captures platform-wide collaborative signals. The joint effort between researchers and engineers at Snapchat resulted in initial UUM models that delivered significant post-launch improvements across multiple applications (e.g., 2.78\% increase in Story Open Rate for Content embedding-based retrieval).
Beyond immediate successes, we outlined promising research directions, such as exploring domain-specific encoders and self-attention with information bottlenecks, which have demonstrated strong offline performance. These advancements underscore UUM's potential to enhance personalization and efficiency across the Snapchat platform, and shed light onto the technical challenges underpinning this space.
% \newpage
% \section{Presenter Bio}
% Clark Mingxuan Ju is a research scientist in the User Model and Personalization (UMaP) research team at Snap Inc., helping building products that improve the user's experiences with Snapchat app leveraging cutting-edge techniques including but not limited to graph machine learning, sequential modeling, and natural language processing. 
% He obtained a Ph.D. degree in Computer Science and Engineering at University of Notre Dame and published at top-tier conferences in machine learning, artificial intelligence, data mining, and natural language processing.
%%
%% The next two lines define the bibliography style to be used, and
%% the bibliography file.
\bibliographystyle{ACM-Reference-Format}
\balance
\bibliography{sample-base}

%%
%% If your work has an appendix, this is the place to put it.
% \newpage
% \appendix

% \section{this is appendix}

% \subsection{Part One}

% % Lorem ipsum dolor sit amet, consectetur adipiscing elit. Morbi
% % malesuada, quam in pulvinar varius, metus nunc fermentum urna, id
% % sollicitudin purus odio sit amet enim. Aliquam ullamcorper eu ipsum
% % vel mollis. Curabitur quis dictum nisl. Phasellus vel semper risus, et
% % lacinia dolor. Integer ultricies commodo sem nec semper.

\end{document}